\documentclass[twocolumn]{article}
\usepackage{setspace}
\usepackage{graphicx}
\usepackage[super]{natbib}
\usepackage{amssymb, latexsym}
\usepackage{amsmath} 
\usepackage{amsthm}
\usepackage{bm}
\usepackage[mathscr]{eucal}
\usepackage{enumerate}
\usepackage{multirow} 
\usepackage[center]{caption}
\usepackage{musicography}
\usepackage{float}

\usepackage{geometry}

\usepackage{authblk}
\usepackage{abstract}

\usepackage{booktabs}
\usepackage{color}
\usepackage{url}

\usepackage{hyperref}

\begin{document}

\title{Computational timbre and tonal system similarity analysis of the music of Northern Myanmar-based Kachin compared to Xinjiang-based Uyghur ethnic groups}

\author{Rolf Bader}
\author{Michael Blaß}
\author{Jonas Franke}
\affil{ Institute of Musicology\\ University of
	Hamburg\\ Neue Rabenstr. 13, 20354 Hamburg, Germany\\
}

\date{\today}

\twocolumn
[
\begin{@twocolumnfalse}
	
	\maketitle
\begin{abstract}
The music of Northern Myanmar Kachin ethnic group is compared to the music of western China, Xijiang based Uyghur music, using timbre and pitch feature extraction and machine learning. Although separated by Tibet, the muqam tradition of Xinjiang might be found in Kachin music due to myths of Kachin origin, as well as linguistic similarities, e.g., the Kachin term 'makan' for a musical piece. Extractions were performed using the apollon and COMSAR (Computational Music and Sound Archiving) frameworks, on which the Ethnographic Sound Recordings Archive (ESRA) is based, using ethnographic recordings from ESRA next to additional pieces. In terms of pitch, tonal systems were compared using Kohonen self-organizing map (SOM), which clearly clusters Kachin and Uyghur musical pieces. This is mainly caused by the Xinjiang muqam music showing just fifth and fourth, while Kachin pieces tend to have a higher fifth and fourth, next to other dissimilarities. Also, the timbre features of spectral centroid and spectral sharpness standard deviation clearly tells Uyghur from Kachin pieces, where Uyghur music shows much larger deviations. Although more features will be compared in the future, like rhythm or melody, these already strong findings might introduce an alternative comparison methodology of ethnic groups beyond traditional linguistic definitions.
\end{abstract}
\end{@twocolumnfalse}
]


\section{Introduction}

The definition of ethnic groups in many countries in Southeast Asia or China has been performed using linguistic terms. In China, e.g., to constitute the VIII. National Congress in 1956 in which all ethnic groups should be represented, anthropologists were sent to field trips determining ethnic groups along common words\cite{Mullaney2011}. Deciding about similarities in ethnic groups in terms of music has been proposed based on musical features, mainly analyzed 'by hand'. Alan Lomax introduced cantrometrics\cite{Lomax1977}, comparing the singing styles of music from all over the world along with a long list of features like homophonic / polyphonic, intense or calm singing, etc. Also, the relation between ethnic groups in terms of music was found closer to gene expression when compared to linguistics\cite{Brown2014}. 

This paper tries to compare the music of two ethnic groups, the northern Myanmar Kachin and the western Chinese Uyghur ethnic groups, using Music Information Retrieval (MIR) extracted features trained in self-organizing Kohonen maps (SOM). The reasons for comparing the two groups are linguistic and political Kachin narratives of a common origin of Kachin with northern ethnic groups, like the Mongols, as collected in several field trips of the first author in these regions over the last about ten years. Still, as the word for ‘song’ in Kachin is makan, a word very close to maqam, or Xinjiang muqam, the music tradition of Central Asia, the comparison with Uyghur pieces is performed first in the paper. Further studies will include a comparison with Mongol music and with regionally much closer music of the large ethnic diversity found in Yunnan, southwestern China.

The paper first describes the music of both groups briefly. After discussing the methodology and the descriptions of the features used, the result section presents dissimilarities, next to similarities between the two groups.

\section{Music of the Kachin}

Kachin State is the most northern state in today’s Myanmar, with Myitkyina as its capital. The word 'Kachin' was mainly coined by the Bama majority in Myanmar, while the Kachin consider themselves as belonging to the ethnic groups of Jinghpaw\cite{Sadan2013}. Still, other groups like the Lisu, Rawang, Lashi, Zai-Wa, or Maru might also consider themselves Kachin when comparing themselves with other regions in Myanmar or with the Bama.

Music is also performed by traditional instruments like the \emph{sum pyi} (bamboo flute) or the \emph{dum ba} (double-reed wind instrument). An important ensemble is the wunpawng band (music band), a marching band playing at weddings, funerals, house opening ceremonies, the Manaw festival, each first Sunday at the local Baptist church, as well as church events like thanksgiving. Plucked and bowed stringed instruments are called \emph{tschibu}. The Jews harp is also played, a tradition found in Southeast Asia for lovers to relate songs with meaning only the lovers know. When playing the melody in public, secret messages could be sent.

Playing Western guitar is very popular among young Kachin. Western instruments can be bought at a local shop in Myitkyina, together with Western electric bass or Rock drum kits. As Rock music in Myanmar is very popular since the 80$^{th}$\cite{MacLachlan2011}, for many in their teens a hope is to become rich and famous, especially among Christian minorities like the Kachin. They often got a solid education in Western music and performing Western musical instruments, mainly guitar, organ, and singing, through Bible schools. 

More contemporary music was raising with DJ and club culture, mainly in Yangon. Experimental music is still very rare in Myanmar\cite{Fermont2016}.

\section{Music of the Uyghur}

The muqam of Uyghur people living in western China Xinjiang Aotonomous Region was traditionally taken as part of the shashmaqam tradition of Central Asia, which was shown to be a canonic construction, where local traditions may deviate strongly\cite{Harris2008}. Still, many musical instruments are clearly of Central Asian origin\cite{Litip2012}. Long-necked plucked instruments include the two-stringed dutar, three-stringed satar, two-necked diltar, or the tambur. Several rawab forms, similar to those known from Usbekistan, are from Dolan, Kashgar, or Kumul tradition, as well as a shepherd and a bass type. There is a barbab, similar to the Chinese pi'pa. Bowed instruments include the ghirjek (in various forms like the kumul, dolan, or ancient, as well as bass-ghirjek), or the bowed qushtar (with bass qushtar). The chang and the qalun are dulcimers. A surnay is used, clearly derived from the Turkish zurna. Flues include a nay and a baliman. Also, a horns called burgha duduk or spaya, a trumpet karnay, or a jews harp qowuz are played. A dap frame drum, a naghara or tabilwas drums next to several percussion instruments are used, like sticks, gongs, cymbals, stones, or wooden spoons.

The tonal system of Uyghur music is mainly described as showing no microtonality, different from many maqam traditions\cite{Harris2008}. Below the comparison of tonal systems is performed, where this is basically confirmed.

Timbre in Uyghur music is reported to be of importance when concatenating sung ghazals, short lyrics mainly lament on the unfulfilled love to God. The ghazal content is not so important as the vowels the lyrics contain, which need to fit the melody sung\cite{Light2008}.

\section{Method}

\subsection{Music collections}

Field recordings of the first author between 2013-2019 in Kachin state were used as the Kachin sample. According to the criteria discussed below, 69 pieces were used. 

The pieces from Kachin region were recorded using many different setups in terms of microphones, recording venues, post-processing ,and the like. Some field recordings in the collection have been done using a four-channel field recorder, with four microphones recordings simultaneously, where in the end the recording was chosen with the least reverberation, broad-band spectrum, and lowest noise. Such recordings, although performed in the forest or jungle, are equivalent to studio recordings due to the high quality of A/D converters and microphones as well as very good microphone positioning. Still, other recordings, especially those during dancing, the gathering of many people, or during the need of strong alcohol consumption by the fieldworker, are of less quality\cite{Lissoir2017}. Sometimes informants or musicians insist on performing in very reverberant spaces, like churches or gathering halls, because they consider the room acoustics there best, although, from an analysis standpoint, this is not the optimum choice as it blurs articulation and dynamics. Indeed there is no objective point for recording such music and in all cases, the most reasonable choice was selected. So, e.g., church music with heavy reverberation is included, as this is what all participants in the ceremony are hearing.

On the other hand, commercial recordings of typical ethnomusicological content often are also not recorded under the best conditions. Sometimes these are the field recordings of ethnomusicologists. Sometimes they are recorded in studios with inferior technical equipment compared to that a contemporary fieldworker carries with him. Many of these recordings are mono or only slightly stereo. Therefore, a four or even eight-track recording with a portable field recorder, placing each microphone close to several instruments, e.g., in a \emph{pat wain}, \emph{gamelan}, or \emph{mohori} orchestra leads to a high source separation already in the recording very useful for further analysis. Even the official muqam recordings used in this investigation distributed via Video-Discs have a very low sound quality which is often heavily distorted.

With commercial Pop, Rock, Hip-Hop, K-Pop, or such recordings, the final mix released seems to be more precise at first, as in the mixing and mastering studio, the sound is manipulated in many ways. Here we can expect the sound to precisely be wanted by the musicians. Indeed, this is again often not the case because of many reasons. The loudness war of heavy compression added during the mastering session is mainly enforced by producers and considered very critical by musicians. Also, most music is streamed as MP3, which is a lossy format. Although with 192 kBit streaming quality with medium quality playback equipment and in medium loudness level, a normal consumer is not able to distinguish it from a lossless format like PCM Wave or FLAC, with increased loudness, the difference is indeed considerable. The use of playback equipment, in general, is also blurring the idea of a commercial recording having a decisive timbre. When mixing a musical track in a recording studio, often two kinds of reference loudspeakers are used, a high-end main monitor and a very low-end stereo setup. The finding is that although a piece might sound fantastic with high-end speakers, it still might sound bad with low-level playback equipment, as cheap in-ears, headphones, car sound systems, or the like. Maybe indeed most music today is heard about in-ears or headphones instead of loudspeakers.

So there is no reference timbre we can rely on. On the other hand, the psychoacoustically relevant features\cite{Schneider2018a}\cite{Schneider2018b} are so robust that we expect them to survive all these changes. A guitar sounds like a guitar, no matter which microphone we use for recording or which playback environment to listen to this recording. Music with changing loudness, roughness, or sharpness over its performance is reflected in all these parameters. Musical pitches, rhythms, etc., are maintained too. To turn it the other way around, if music is recorded or presented in a way it does no longer meet these parameters, we would no longer consider them as valid recordings or presentations. This can be judged by ears, and therefore recordings not suitable for a study have been excluded.

Although the first author also did fieldwork in Xinjiang among the Uyghur people, not enough recordings were collected to have a reasonable comparison. Therefore the Uyghur pieces were collected from these recordings:

\begin{itemize}
	\item Uighur Autonomous Region Song and Dance Ensemble, King Records, 1991 (instruments: rewap, balaman, dutar, tambur, dombura, tambourine, voice, ghijak, husitar, rawap, satar, ajek, dap)
	\item Sanubar Tursun, Felmay, Italy, 2013 (instruments: Voice, tambur, satar, dutar).
	\item Collected by Laurent Jeanneau, Shi Tanding, Subleme Frequencies, 2009 (instruments: dongbra, tamburg, voice, dotar, satar, ,omuz, rawa, topchar, tschang)
	\item Qetik, Taklamakan Rock, Dreyer Gaido Musikproduktionen, 2013 (instruments: Vocals, guitar, dumbura, keyboards, drums, bass).
	\item muqäddimä of 12 muqam of the China Uighur Twelve Muqam Symposium\cite{UyghurSymposium} recordings of the Uighur Autonomous Region, Song and Dance Ensemble.
\end{itemize} 

The Uighur Autonomous Region Song and Dance Ensemble was the official state ensemble located in Ürümqi. The recordings of Sanubar Tursun are mainly voice and a single instrument. The collection of Laurent Jeanneau and Shi Tanding include various artist. The Qetik recording is Rock music with many traditional components. The 12 muqäddimä are taken from the official release of the Uighur Song and Dance Ensemble, which covers twelve CDs (VCDs) of the canonical 12 muqam. From each the introduction section, the muqäddimä, was taken which is mostly solo voice accompanied by a single instrument. Altogether 63 Uyghur pieces were used.

To arrive at a more widespread field in terms of ethnic groups of interest, the two ethnic groups are framed within a larger sample of songs from Southeast Asia, China, Sri Lanka, and Nepal. Still, as the scope of this paper is to compare Kachin and Uyghur pieces, the results concentrate on their relations only.

\subsection{Feature Extraction and Machine Learning}

The musical pieces are analyzed with respect to timbre and tonal system. The analysis is two-step. 

\begin{enumerate}
	 \item In a first step, using algorithms of Music Information Retrieval (MIR), features are extracted. 
	 \item In a secon step, using a Kohonen self-organizing map (SOM), the extracted feature vectors arrange in a two-dimensional neural field, displaying clusters.
\end{enumerate}

Investigating what is different in the music of ethnic groups does include determining what is similar. All methods used show both the similar and the dissimilar. Comparing different musical styles can therefore be performed qualitatively, as well as quantitatively. Ending in one judgment how far away the music of two ethnic groups is is not the aim of this paper. It is rather to give insights into details. Such a judgment would include a decision which features to take for such a judgment, how to weight them, and how to combine them to arrive at a single value. If this is reasonable to do might be discussed elsewhere.

\subsubsection{Timbre similarity}

There are two kinds of analysis, a forward and a backward or inverse analysis. In the forward analysis, musical pieces are analyzed with features chosen from a feature set and sorted in the SOM. There are two decisions to make, a) which musical pieces to analyze, and b) which features to use. Depending on these choices, the SOM similarity estimation is different.

The inverse or backward analysis is going the same way back\footnote{For a systematic description of Inverse Problems, see \cite{Groetsch1993}.}. Starting from a research question with an intended SOM result, like separating musical pieces in a two-dimensional map, those pieces and audio features are chosen, which arrive at this estimation. Still, there is no strict proof that this is a global maximum, the optimum choice. Still, if we arrive at a clustering, we have found relevant psychoacoustic parameters separating the musical corpora. 

In this paper, the reasoning for inverse solving is that of psychoacoustics. Within the last about 150 years, starting from Helmholtz’s idea of perceptual roughness as the basis for musical scale \cite{Helmholtz1863}, psychoacoustic research has extensively been performed, and most salient timbre features perceived by listeners all over the world, and respective analyzing MIR algorithms are known. Discussing these parameters in detail is beyond the scope of this paper, for reviews see \cite{Schneider2018a}\cite{Schneider2018b}\cite{Bader2013}.

\subsubsection{Tonal system similarity}

Tonal systems are most often understood as a small set of cent values, pitches of a musical piece, located within one octave. Still, performance practice leads to a variety of pitches used for played notes. Therefore, in this study, tonal systems are taken as histograms of all pitches performed in musical notes, accumulated in 1 cent intervals within one octave, resulting in 1200 cent values. This definition includes all deviations, vibrato, melismas to a certain extend, or different pitches presented at different occurrences of a note. As shown in the result section, this approach clusters the music investigated in this study.

\subsection{Music Information Retrieval (MIR) analysis}

The analysis is performed with the apollon\cite{apollon} framework of the Computational Music and Sound Archiving (COMSAR)\cite{COMSAR} project at the Institute of Systematic Musicology at the University of Hamburg\cite{Bader2019a}\cite{COMSAR}. It is implemented as an online version in the Ethnographic Sound Recordings Archive (ESRA)\cite{ESRA} of historical and contemporary music ethnological recordings at the Institute (The 1932 Cairo Congress of Arab Music, Collectin Wilhelm Heinitz of African Music, Collection Bader). It can also be used as an offline tool in a jupyter notebook browser environment based on Python programming language.

In apollon and COMSAR many analysis tools for timbre, rhythm, tonal system, or melody are implemented. All of them take the digital or digitalized recordings as input. We restrict our presentation to only those features used in the present paper, i.e., melody, tonal system, and timbre.

\subsubsection{$f_0$ estimation and note detection}

\begin{figure}
	\centering
	\includegraphics[width=1\linewidth]{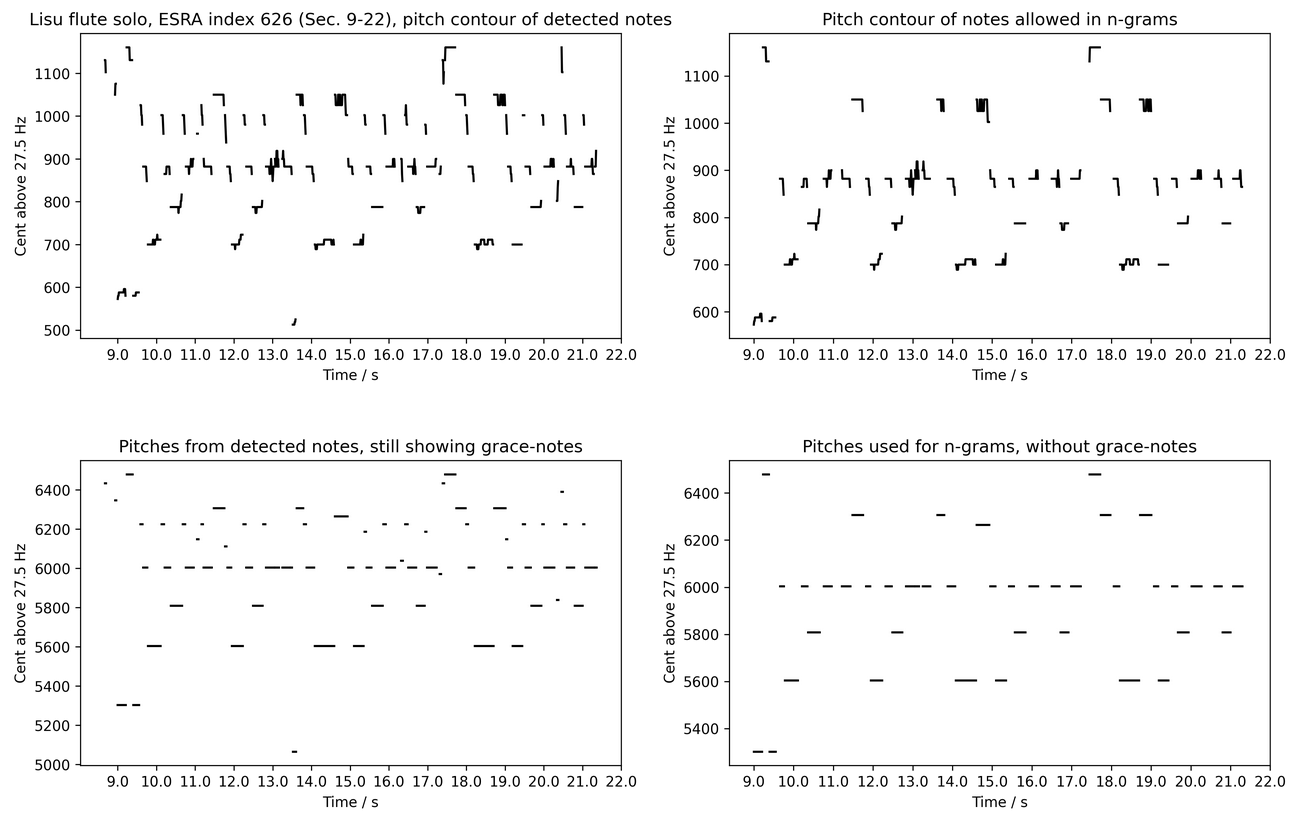}
	\caption{Example of pitch and melody extraction using Lisu flute solo, ESRA index 626\url{https://esra.fbkultur.uni-hamburg.de/explore/view?entity_id=626}. The analysis has five stages of abstraction. 1) $f_0$ is calculated over the whole piece. From $f_0$ values to melodies: 2) top left: Pitch contours of detected notes, 3) top right) pitch contour of notes allowed for n-grams (melodies), 4) bottom left: mean pitches of allowed notes from plot 2) still showing grace-notes, 5) bottom right: mean pitches of notes allowed for n-grams.}
	\label{fig:pitchmelodylisu626}
\end{figure}

All pieces are analyzed in terms of pitch. Five levels of abstraction are performed, as shown for the example of a Lisu solo flute piece in \textbf{Figure} \ref{fig:pitchmelodylisu626} \footnote{ESRA index 626\url{https://esra.fbkultur.uni-hamburg.de/explore/view?entity_id=626}, \url{https://vimeo.com/showcase/5259277/video/278497941}}.

In a first analysis stage, $f_0$ extraction is performed, with overlapping frames of 25 ms resulting in 100 $f_0$ values per second using an autocorrelation algorithm\cite{Six2013}\cite{Mauch2014}, allowing a minimum frequency of 40 Hz. The $f_0$ values are transferred into cent values, starting at $f_{min} = 27.5$ Hz, the subcontra A and allowing eight octaves above $f_{min}$, which is more a theoretical choice as higher octaves are way above pitches used in traditional music. 

The second abstraction stage uses an agent-based approach, where musical events, notes, grace-notes, slurs, melismas, etc., are detected. The agent follows the cent values from the start of each musical piece and concatenates adjacent cent values according to two constraints, a minimum length a pitch event needs to have, here 30 ms, and a maximum allowed cent deviation, here $\pm$ 60 cents. This allows for including vibrato and pitch glides within about one semitone, often found in vocal and some instrumental music. With flues, often very straight pitches were found. Still, these pitches might be different at different parts of the piece. Lowering the allowed deviation leads to to the exclusion of pitches that often have quite strong deviations. As an example, an excerpt of 13 seconds of a Lisu solo flute piece is shown in \textbf{Figure} \ref{fig:pitchmelodylisu626} on the top left. Some pitches show a quite regular periodicity, some are slurs or grace-notes.

The third abstraction stage determines single pitches for each detected event by taking the strongest value of a pitch histogram. As can be seen in the top left \textbf{Figure}, pitches are often stable, only to end in some slur in the end. Therefore, taking the mean of these pitches would not represent the main pitch. Using the maximum of a histogram, on the other side, detects the pitch most frequency occurring during the event. On the bottom left, this is performed and can be compared to the top left plot. When listening to the piece, this representation seems to contain still too many pitch events. So. e.g., the events around 6000 cents (above $f_{min}$ = 27.5 Hz) are clearly perceived as notes. Still, those small events preceding around 6200 cents are heard as grace-notes. Therefore, to obtain a melody without grace-notes, a fourth stage needs to be performed.

In this fourth stage, pitch events are selected using three constraints to qualify as melody notes allow for n-gram construction.  n-grams have shown to represent melodies stable and robust in terms of melody identification\cite{Downie1999}. Still, no n-gram melody SOM is constructed, as detecting melodies from polyphonic pieces is not covered by autocorrelation algorithms. The amount of single-line pieces in the Kachin music collection was too small to allow comparison with other ethnic group music. Still, note detection needs to be performed detecting tonal systems, as the pitch values calculated by autocorrelation also include transients, pauses, or other non-pitch events.

To qualify as a note, excluding grace-notes, slurs, etc., the notes need to have a minimum length, here 100 ms. Additionally, a lower and upper limit for adjacent note intervals is applied for n-gram construction. The lower limit is 0 cent here to allow tone repetition, especially expected for Uyghur pieces. The upper limit was set to $\pm$ 1200 cent, so two octaves, most often enough for traditional music. This does not mean that traditional pieces do not have larger intervals, as, e.g., expected in jodeling. Still, such techniques are not used in the present music corpus, and even when present, they are not expected to be more frequent than smaller intervals. Indeed, small intervals are most prominent in traditional music, as also confirmed in the present study. In \textbf{Figure} \ref{fig:pitchmelodylisu626} top right, we see the pitch contours for all allowed notes used in n-gram calculation. Indeed, all grace-notes are gone.

In the last step, shown in the bottom right plot, the pitches of each event are again taken as the maximum of the historgram of each event. Now following the musical piece aurally, the events represent the melody.

\subsubsection{Tonal System}

Since the proposal to calculate tonal systems using pitch and cent values\cite{Ellis1884} this method was widely used to determine the tuning of musical instruments. Gamelan tunings in gong kebyar tended to sharper intervals over the course of the 20$^{th}$ century\cite{Tenzer2000}. In listening tests, it was found that by varying the tuning of gamelan, the original tuning was considered most suitable, although iz shows a maximum roughness\cite{Wendt2019}. Historical development of tuning systems was shown for Burmese music, including so-called neural intervals also influenced by Thai music\cite{Williamson2000}. In a survey of tuning systems, a preferred interval of around 170 cents was found for tonal systems in Southeast Asia, which corresponds to the interval of a 7-tone equidistant scale\cite{Mcbride2019}. Automatic Indian Rāga estimation has been performed based on pitch estimation\cite{Koduro2012}\cite{Chordia2013}. Taking musical instrument building and performance practice into consideration, it was argued that tonal systems in Southeast Asia are temperaments, just like in Western music, compromises between musical instrument building, performance practice, and harmonic intervals\cite{Bader2019b}. 

A tonal system in this study, therefore, is the accumulation of all $f_0$ values transferred into cent of all notes, extracted as discussed above, within one octave. The lowest frequency of this octave is the peak of all $f_0$ values accumulated over eight octaves, starting from $f_{min} = 27.5$ Hz. A precision of 1 cent is used to arrive at a tonal system vector of 1200 entries.

Contrary to melody extraction, which can only be reasonably performed for single-line melodies, tonal systems can also be extracted from some musical pieces where, e.g.,  a singer is accompanied by a dutar, or a strong melody instrument is overwhelming an ensemble. Although when a real polyphonic texture is present, the algorithm fails, for some part of the piece it calculates performed pitches. Therefore, in the tonal system analysis, those pieces which show correct melodic fragments were included in the sample. The note identification algorithm discussed above ensures that only these correct pitches are used in the tonal system analysis. If a piece showed too few such cases, it was excluded.

\subsubsection{Timbre}

All timbre features use a Fourier Transform to analyze the sound in adjacent, non-overlapping frames of 80 ms. The analyzed discrete spectrum $A_i$ with amplitude A and N frequency entries $f_i$ is then further processed in using four features. Here i is used as the vector bins, which map into frequency values through the sample frequency s and the analysis window length l in samples like $f_i$ = i / (l / s). The choice of the parameters is motivated by psychoacoustics, as discussed above.

\textbf{Spectral Centroid}

The spectral centroid C is the center of a spectrum, where the sum of amplitudes of frequencies above and below this center are equal, and is calculated as

\begin{align}
C = \frac{\sum_{i=0}^N f_i A_i}{\sum_{i=0}^N A_i} \ .
\end{align}

This corresponds to psychoacoustic brightness perception.

\textbf{Roughness}

Roughness calculations have been suggested in several ways (for a review, see \cite{Schneider2009}, \cite{Bader2013}.). Basically, two algorithms exist, calculating the beating of two sinusoidals close to each other (Helmholtz\cite{Helmholtz1863}, Helmholtz/Bader\cite{Schneider2009}, Sethares\cite{Sethares1993}), or integrating energy in critical bands on the cochlear (Fastl\cite{Zwicker1999}, Sottek\cite{Sottek1994}). The former has been found to work very well with musical sounds, the latter with industrial noise. 

In this paper, a modified Helmholtz/Bader algorithm is used. Like Helmholtz, it assumes a maximum roughness of two sinusoidals at 33 Hz frequency difference. As Helmholtz did not give a mathematical formula how he did calculate roughness, according to his verbal descriptions, a curve of the amount of roughness $R_n$ is assumed between two frequencies with distance $df_n$ which have amplitudes $A_1$ and $A_2$ like

\begin{equation}
R_n = A_1 A_2 \frac{|df_n|}{f_r e^{-1}} e^{- |df_n|/f_r} \ .
\end{equation}

with a maximum roughness at $f_r$ = 33 Hz. The roughness R is calculated as the sum of all possible sinusoidal combinations like

\begin{equation}
R = \sum_{i=1}^N R_i \ .
\end{equation}

The only difference between the algorithm used in apollon and that described in \cite{Schneider2009} is the precision with which the frequencies are calculated. To arrive at very precise values in \cite{Schneider2009} a wavelet analysis is performed, allowing for an arbitrary precision of frequency estimation. As this is very expensive in terms of computational time, in the present study, the above-described Fourier analysis precision is used. In \cite{Schneider2009} the research aim was to tell the perceptual differences between tuning systems like Pure Tone, Werkmeister, Kirnberger, etc. in a Baroque piece of J.S. Bach. The present analysis is not aiming for such subtle differences but for the overall estimation of roughness.

\textbf{Sharpness}

Perceptual sharpness is related to the work of Bismarck\cite{Bismarck1974} and followers \cite{Aures1985a}\cite{Aures1985b}\cite{Fastl2007}. It corresponds to small frequency-band energy. According to \cite{Fastl2007} it is measured in acum, where 1 acum is a small-band noise within one critical band around 1 kHz at 60 dB loudness level. Sharpness increases with frequency in a nonlinear way. If a small-band noise increases its center frequency from about 200 Hz to 3 kHz sharpness increases slightly, but above 3 kHz strongly, according to perception that very high small-band sounds have strong sharpness. Still, sharpness is mostly independent of overall loudness, spectral centroid, or roughness, and therefore qualifies as a parameter on its own.

To calculate sharpness, the spectrum A is integrated with respect to 24 critical or Bark bands, as we are considering small-band noise. With loudness $L_B$ at each Bark band B sharpness is

\begin{equation}
S = 0.11 \frac{\sum_{B=0}^{24 Bark} L_B g_B B}{\sum_{B=0}^{24 Bark} L_B} \ \text{acum} ,  
\end{equation} 

where a weighting function $g_B$ is used strengthening sharpness above 3 kHz like\cite{Peeters2004}

\begin{equation}
g_B = \left\{\begin{array}{ll} 1 \text{ if} B < 15 \\ 0.066 e^{0.171 B} \text{ if} z \geq 15 \end{array} \right.
\end{equation}

\textbf{Loudness}

Although several algorithms of sound loudness have been proposed\cite{Fastl2007}, for music, still no satisfying results have been obtained\cite{Ruschkowski2013}. Most loudness algorithms aim for industrial noise, and it appears that musical content considerably contributes to perceived loudness. Also, loudness is found to statistically significantly differ between male and female subjects due to the different constructions of the outer ears between the sexes. Therefore a very simple estimation of loudness is used, and further investigations in the subject are needed. The algorithm used is

\begin{equation}
L = 20 \log_{10} \frac{1}{N}\sqrt{\sum_{i=0}^N \frac{A_i^2}{A_{ref}^2}} \ .
\end{equation}

This corresponds to the definition of decibel, using a rough logarithm-of-ten compression according to perception and multiplying with 20 to arrive at 120 dB for a sound pressure level of about 1 Pa. Of course, the digital audio data are not physical sound pressure levels (SPL). Still, the algorithm is used to obtain dB-values most readers are used to. As all psychoacoustic parameters are normalized before inputting them into the SOM, the absolute value is not relevant.

All timbre parameters were integrated over the whole piece as mean and standard deviation. Concatenating them arrives at the feature SOM training vector.

\subsubsection{Machine learning using self-organizing neural maps (SOMS)}

The COMSAR architecture used in this paper uses the self-organizing map\cite{Kohonen2001}. The two-dimensional map consists of 26 $\times$ 26 neurons for the tonal system and 15 $\times$ 15 for timbre feature vectors. As the timbre feature vector is much smaller than that of tonal systems, for tonal systems, a larger neural net allows for a better differentiation for the larger vector. The training set consists of either the timbre feature vector or the tonal system feature vector of all pieces in the collections, including Yunnan, Sri Lanka, Nepal, and related regions. Training showed fast convergence in both cases, and 500 iteration steps were used each.
Training the map means correlating the feature vector of each musical piece with all feature vectors of the neural map, which is initiated randomly at the beginning. For each trained feature vector, the neuron on the map with the strongest correlation and its neighboring neurons are altered towards the training vector using a Mexican-hat function. This is repeated for all training vectors, so for all musical pieces, the map is trained with. After one such cycle, the map already starts self-organizing, where distinct training feature vectors tune different regions of the map towards the training data. This training is repeated in 500 cycles for each training set.

After training, different regions of the map are similar within the region and slide over to other regions more dissimilar. A u-matrix shows the similarity or dissimilarity of neighboring neurons on the map to display where regions of similarity are present. Therefore, the distance between two neurons is not only the spatial distance on the map but also displayed by the similarity between neighboring neurons. In the plots below, blue and green regions show strong similarity, while brown, gray, or white regions are very dissimilar.

Arbitrary feature vectors, musical pieces, can be positioned on the trained map, where their position is the neuron with the highest correlation with respect to the musical piece. The distribution of musical pieces from different ethnic groups shown below therefore tells if the music of these groups can be identified and distinguished by a feature, here timbre or tonal system, or if the group’s music is similar in this respect.

\section{Results}

\subsection{Tonal system Kachin/Uyghur comparison}

\begin{figure}
	\centering
	\includegraphics[width=1\linewidth]{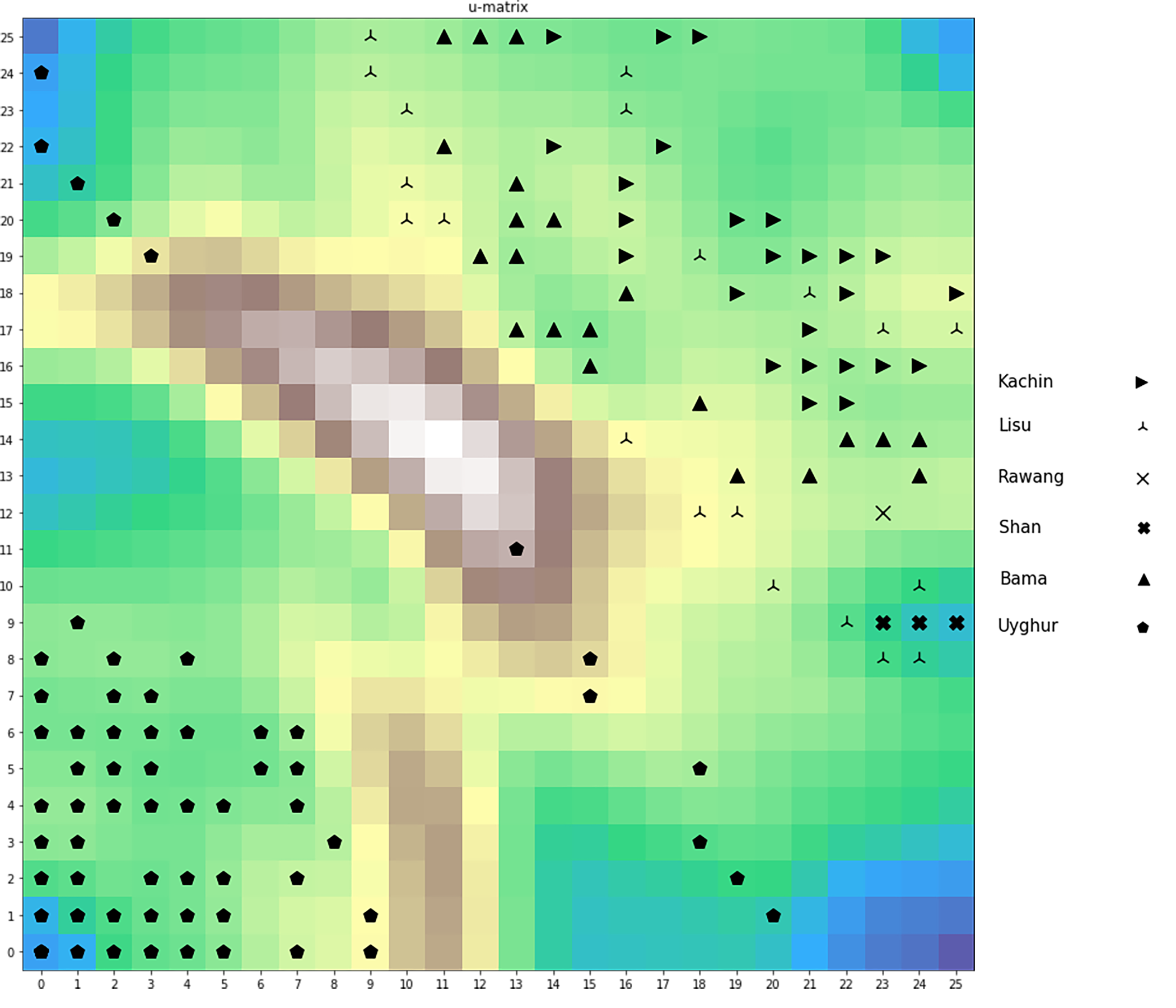}
	\caption{The Kohonen neural map trained with collection Bader and Uyghur musical pieces in terms of their tonal systems. The background color is the u-matrix, telling how similar or dissimilar neighboring neurons are, where blue and green denote similar, brown, gray, or white denote dissimilar neighborhoods. Therefore, the plot shows a strong split between Uyghur pieces, mainly on the lower left, and Lisu, Kachin, Rawang, Shan, or Bama pieces on the upper right side.}
	\label{fig:tonalsystemuyghurlisurawangshanbama}
\end{figure}

\textbf{Figure} \ref{fig:tonalsystemuyghurlisurawangshanbama} shows the u-matrix of the SOM trained with the collection Bader and other Uyghur recordings with respect to their tonal system. As described above, each neuron is a feature vector of length 1200, containing the accumulated cent values for all notes within each piece. The background color displaying the similarity between neighboring neurons shows a clear ridge, splitting the lower-left corner from the upper right side. Indeed, most Uyghur pieces are in the lower-left corner, with additional pieces on the upper left corner and some following the main ridge downwards. All Kachin, Lisu, Rawang, Shan, and Bama pieces, also shown in the map, are on the right upper corner. Therefore, a clear difference between the tonal system used by the Uyghur and the Kachin and related ethnic group’s music appears.

\begin{figure}
	\centering
	\includegraphics[width=1\linewidth]{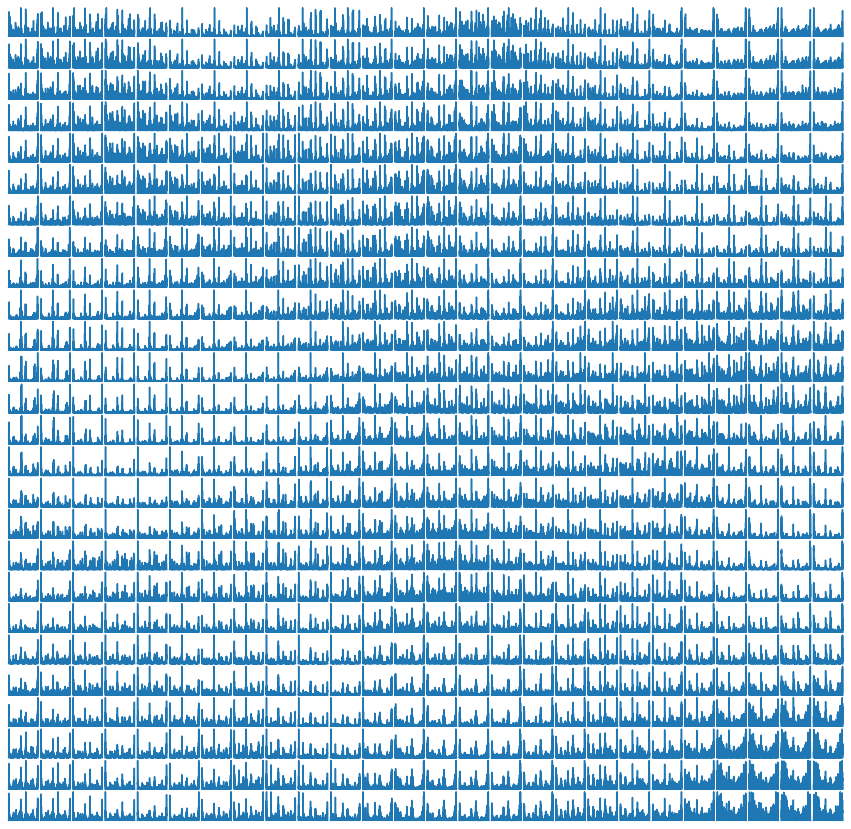}
	\caption{Tonal systems as trained by the map in \textbf{Figure} \ref{fig:tonalsystemuyghurlisurawangshanbama}, showing regions of similar and dissimilar tonal systems.}
	\label{fig:tonalsystemallcents}
\end{figure}

To have an idea about the regions of similar tonal systems, in \textbf{Figure} \ref{fig:tonalsystemallcents} the tonal systems of all neurons on the map are shown. Comparing with \textbf{Figure} \ref{fig:tonalsystemuyghurlisurawangshanbama}, the lower-left corner where many Uyghur pieces are located shows distinguished pitches, while other regions on the right upper side have a more widespread distribution of pitches over the octave.

\begin{figure}
	\centering
	\includegraphics[width=1\linewidth]{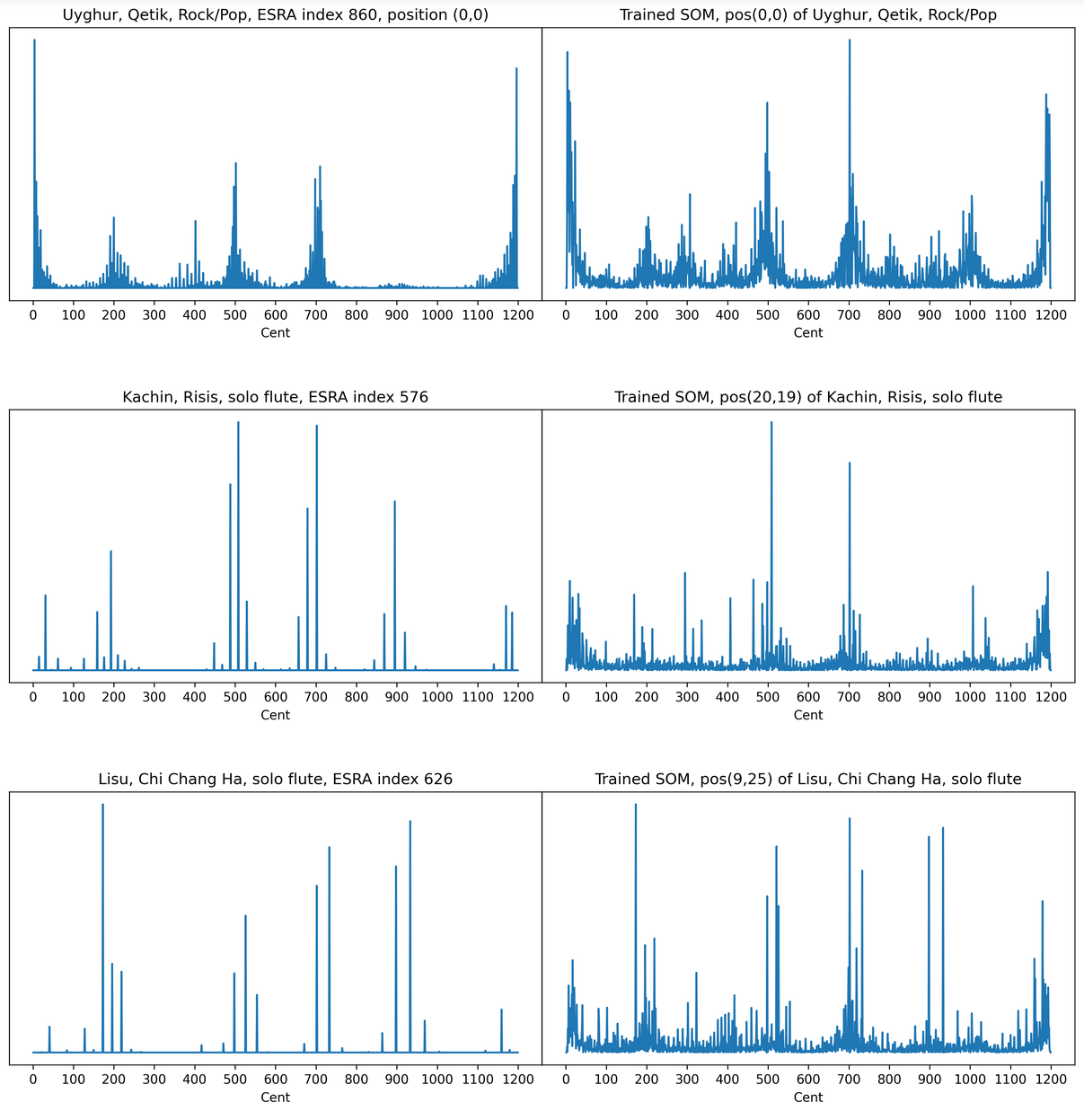}
	\caption{Three examples of tonal systems as calculated from sound file (left column) and as vector on the neural map on the location the musical pieces fits best (right column). Top: Uygur Rock/Pop piece by Qetik, Middel: Kachin flute solo piece, Bottom: Lisu flute solo piece.}
	\label{fig:tonalsystemexamples}
\end{figure}

Multiple musical pieces can be placed on a single neuron. In most cases, on the map, only one or two pieces are present. Still, on position (0,0), 11 pieces are located, which are all from the Uyghur Rock/Pop band Qetik. Other pieces around position (0,0) are traditional pieces, like those of Sanubar Tursun, the Xinjiang muqam art ensemble, or other such recordings. The Qetik tracks are multi-track. Still, analysis is possible here due to strong melody lines in the tracks detected by the algorithm. In \textbf{Figure} \ref{fig:tonalsystemexamples} in the first row, one example of the Qetik piece, Derdi Tolidur Yarning is shown. On the left, the analysis result of the piece. On the right, the map feature vector, the trained tonal system on position (0,0) is shown where the piece is located. Remember that the tonal system map on the right is a representation of all musical piece’s tonal systems located on this node. Although the tonal system on the left has no perfectly clear peaks, the trained tonal system has. Its fifth, e.g., is at 702 cent, representing just intonation rather than an equidistant 12-tone scale. Still, the fifth is at 503 cent, 498 cent would be just intonation, and a minor third at 307 cent, again 316 would be just intonation. Still, as all Qetik pieces are located at this node, the tonal system map at (0,0) is closest to all these pieces compared to all other map vectors. As shown in \textbf{Figure} \ref{fig:tonalsystemallcents}, neighboring nodes around (0,0) are similar, and therefore the vector shown in \textbf{Figure} \ref{fig:tonalsystemexamples} for (0,0) is like a fingerprint to Uyghur musical pieces tonal systems analyzed in this study.

In \textbf{Figure}\ref{fig:tonalsystemexamples} in the middle row, a Kachin solo flute piece is shown. The recording analysis shows sharp peaks, still distributed around fifth and fourth, next to the major second. The trained SOM vector this piece fits best finds these fifths and fourths to be most characteristic for this piece, always in comparison to all other pieces the map is trained by.

Yet a third example is again a flute solo piece of Lisu music showing considerably stronger pitches around the sixth (around 900 cent) and strong peaks around the major second (around 200 cent), while fifth and fourth are not so prominent compared to the piece in the middle row of the \textbf{Figure}. The trained map shows the sixth and the second very prominent, while also strong fifth and fourth are present. Both these trained SOM tonal systems are therefore fingerprints of the musical pieces, not only in terms of pitches present but also according to how often these pitches appear in the pieces.

\begin{figure}
	\centering
	\includegraphics[width=1\linewidth]{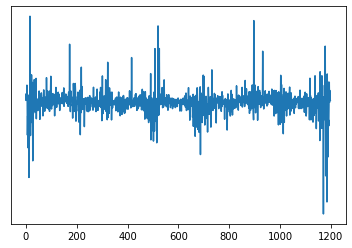}
	\caption{Difference of mean tonal systems of upper right and lower left triangular matrix of the tonal system map in \textbf{Figure} \ref{fig:tonalsystemuyghurlisurawangshanbama}, showing enhanced or reduced tone regions for Kachin and associated ethnic groups vs. Uyghur pieces.}
	\label{fig:tonalsystemuyghurkachindifference}
\end{figure}

\begin{figure}
	\centering
	\includegraphics[width=1\linewidth]{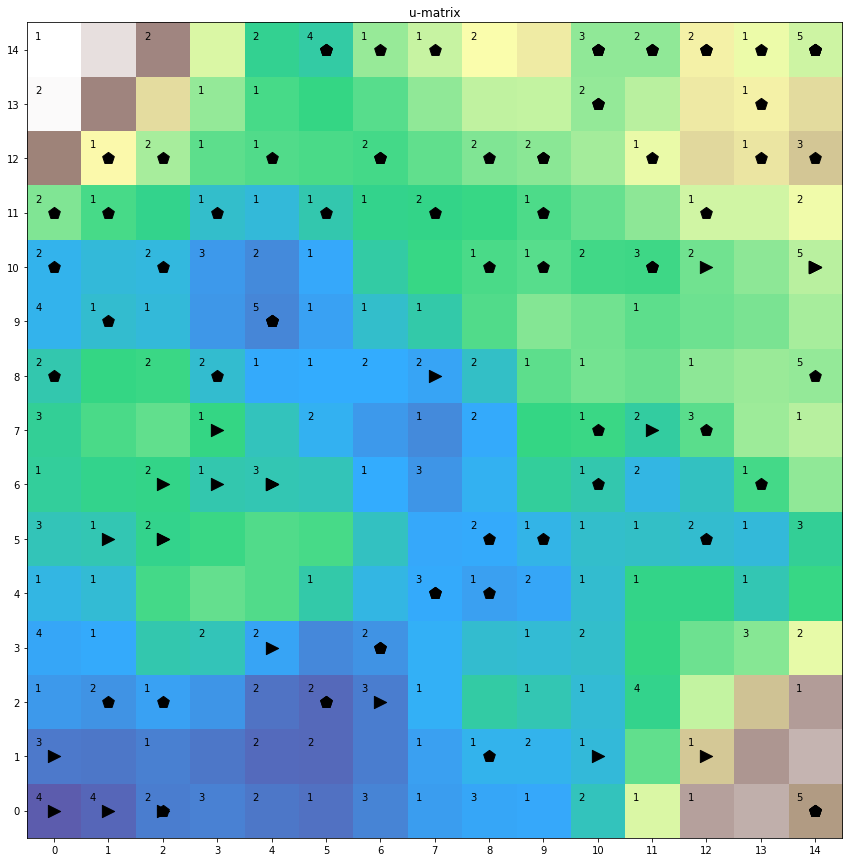}
	\caption{Self-organized map (SOM) trained using timbre feature vectors extracted by MIR tools of Collection Bader (ESRA Archive) and Uyghur pieces, with best-matches for Uyghur musical pieces (Xinjiang, China), and Kachin pieces 
		Northern Myanmar). Colors indicate the similarity of neighboring neurons (brown and green are more dissimilar, blue more similar). Most Uyghur pieces are found in the upper half with more dissimilarity, while Kachin pieces appear to be more similar.}
	
	\label{fig:somumatrixuyghurkachin}
\end{figure}

Therefore we can conclude that Uyghur and Kachin musical pieces clearly have different tonal systems. They are split by a sharp ridge in the u-matrix. Also pieces within Kachin music are present, differentiating Kachin, Lisu, or Rawang pieces.

To estimate main differences, the map is split into an upper right and a lower left triangular matrix, justified by the clear ridge in the u-matrix. The upper right side contains the Kachin and related ethnic groups, while the lower left is the Uyghur side. For both triangular matrices, a mean tonal system is calculated, and the lower-left mean is subtracted from the upper right. \textbf{Figure} \ref{fig:tonalsystemuyghurkachindifference} shows this difference. Peaks larger than zero mean tone regions more prominent with Kachin, and peaks below zero tone regions more prominent in Uyghur tonal systems. The lowest, fundamental (0 cent) is omitted, as it outperforms all other peaks in favor of Kachin to make the rest of the plot better readable. The fifth region shows that Kachin pieces tend to have a higher fifth compared to Uyghur pieces. The same holds for the fourth. In both cases, Uyghurs center around just intervals, while Kachin tends to be higher. There are sharp peaks at the major sixth, major second, and major third regions stronger with Kachin pieces, and an enhanced seventh region with the Uyghur music. Still, the dominance of a fundamental at 0 cent with Kachin music compared to Uyghur pieces outperforms all these minor differences. The standard deviation of both regions is nearly perfectly the same, pointing to a similar consistence of tonal systems between the two groups.

Note that this definition of tonal system deviates from that of a small set of cent values, one for each note in the tonal system. As recordings nearly never show such clear pitches, the suggested method is that of correlating an amplitude spectrum of pitches present in musical pieces. These are fingerprints with characteristics such as presence, slight presence, or absence of pitches, deviations of pitches from a maximum, or bright vs. sharp spectra. All these contribute to the perception of a tonal system present in musical pieces.

\subsection{Timbre-based Kachin/Uyghur comparison}

\textbf{Figure} \ref{fig:somumatrixuyghurkachin} shows a self-organizing map trained by Collection Bader pieces of Southeast Asia, China, Sri Lanka, or Nepal. For the training vector, timbre features have been extracted from the pieces using MIR tools. After training, music from Uyghur ethnic group (Xinjiang, China), as well as Kachin ethnic group (Northern Myanmar) have been plotted as the best fit to the 15 x 15 neuron matrix. The background color shows the similarity between neighboring neurons, where white, brown, or green mean more dissimilarity, and blue indicates stronger similarity. 

\begin{figure}[!]
	\centering
	\includegraphics[width=1\linewidth]{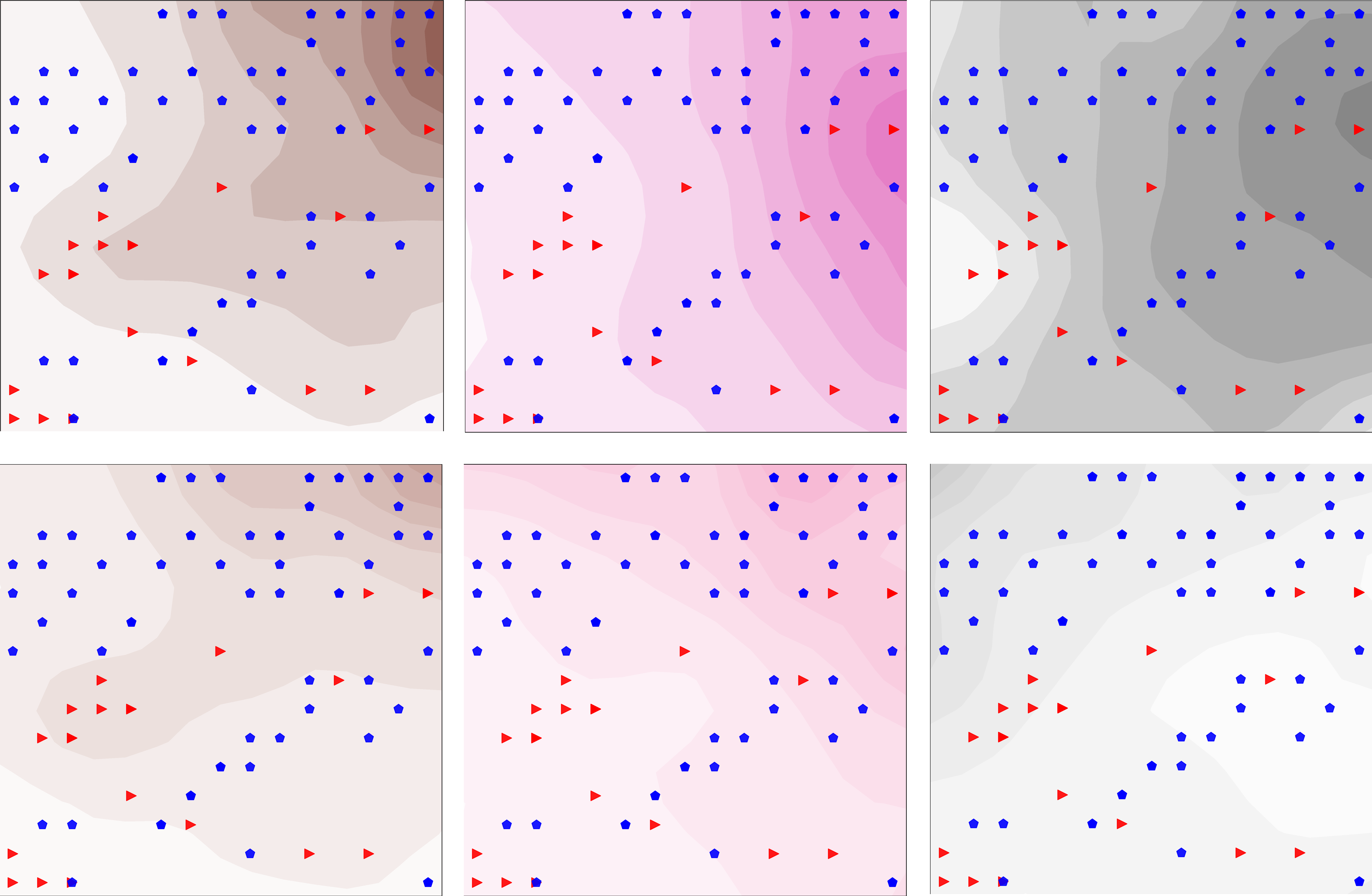}
	\caption{Correlation of trained SOM feature vector with roughness (left), sharpness (middle), and loudness (left) for mean (upper plots) and standard deviation (lower plots). So e.g. the neurons in the upper right corner have the highest mean and standard deviation of roughness. Pieces with best-fit in this region therefore are also strong in these parameters.}
	\label{fig:somkachinuyghurroughsharploudmeansd}
\end{figure}

In the upper third of the plot, only Uyghur pieces are found, correlating with roughness mean and standard deviations (upper left corner) and sharpness standard deviation (upper middle) (see \textbf{Figure} \ref{fig:somkachinuyghurroughsharploudmeansd})). Musically the high standard deviation corresponds to a strong fluctuation of these parameters within the pieces.

A correlation between traditional and modern music is found at the best-fit pieces on the neuron with the highest sharpness standard deviation, where a traditional piece Waderiha, as well as the Qetin Rock piece ‘Dolan muqam’ are both located. This shows this characteristic timbre feature of strong sharpness variations during the piece to be continued in Rock music of modern Uyghur performers.

Closer examination of the Uyghur pieces at the upper left side shows correspondence with high centroid and loudness standard deviation and overall low loudness. All these pieces have a characteristic bel canto-like singing style, where all such pieces are only found in this plot region.

The Kachin pieces are most often sorted in blue regions of \textbf{Figure} \ref{fig:somumatrixuyghurkachin}, indicating a stronger similarity within the music of this ethnic group. One exception is the Kachin Wunpawng marching band, consisting of drums and zurna-like instruments, which have their best-fit at the position of highest sharpness and loudness neurons (\textbf{Figure} \ref{fig:somkachinuyghurroughsharploudmeansd}, maximum at upper right plot). Indeed, perceptually this music is very loud. 

\subsection{Detailed display of features}

\begin{figure}
	\centering
	\includegraphics[width=1\linewidth]{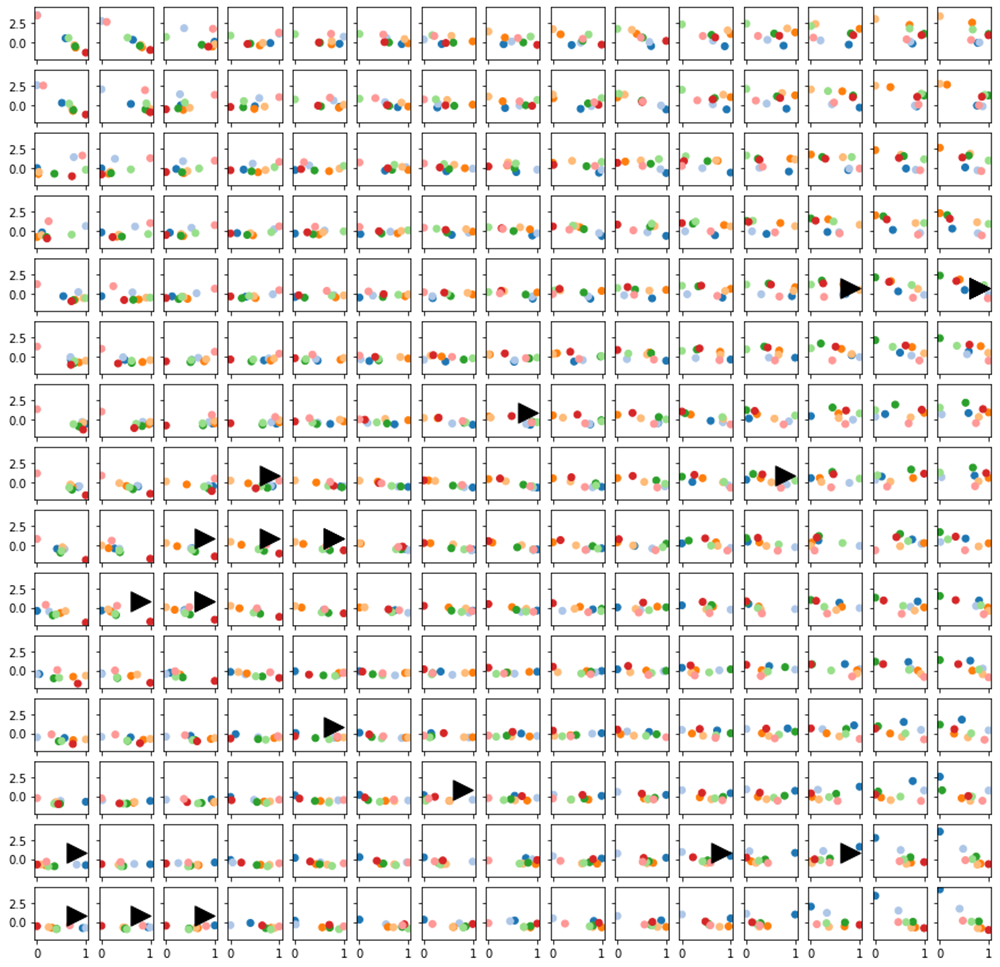}
	\caption{Detailed display of feature strength and feature importance with respect to sorting a musical piece at a neuron position for the Kachin pieces. The up-down axis shows the strength of each feature, the left-right axis its importance with respect to sorting a piece at this position, where the most left features are the most important. Spectral centroid (blue), Roughness (orange), Sharpness (green), Loudness (red), mean is darker color, standard deviation is brighter color (e.g. dark blue = mean centroid, light blue = standard deviation centroid).}
	\label{fig:somkachinfeatureimportance}
\end{figure}

\begin{figure}
	\centering
	\includegraphics[width=1\linewidth]{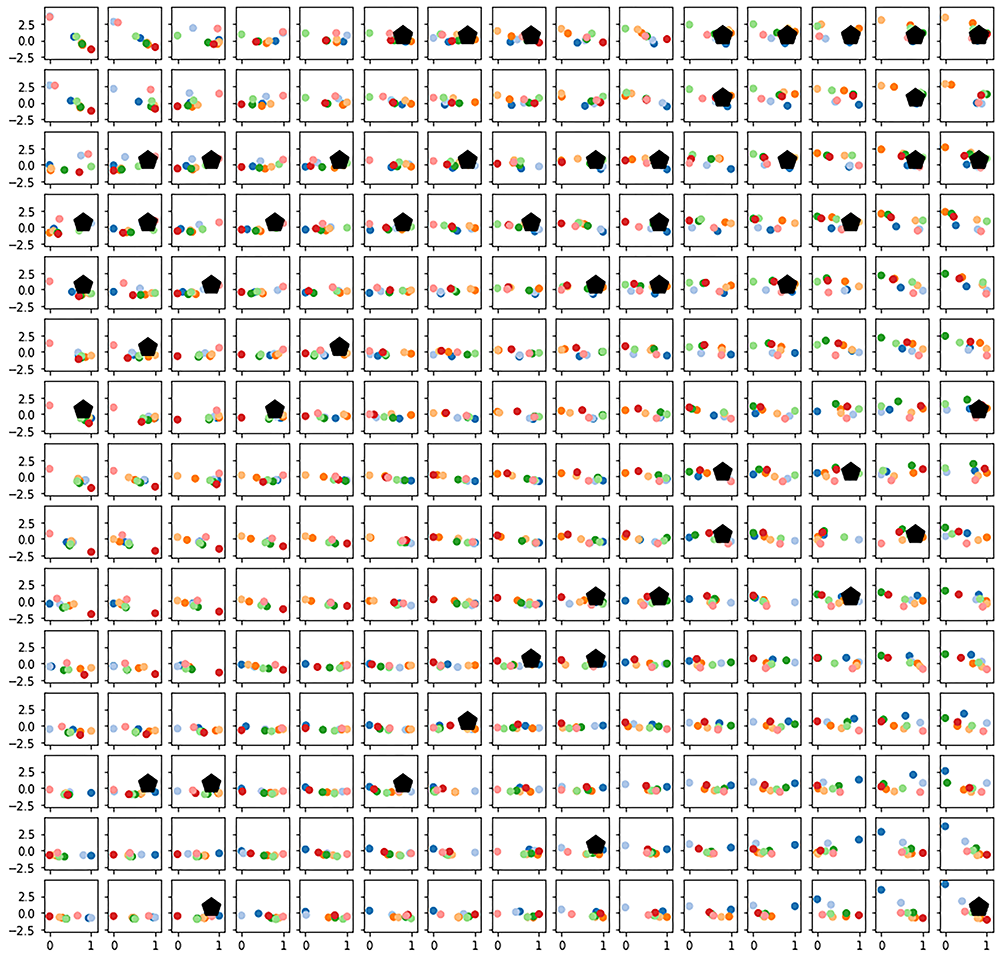}
	\caption{Same as \ref{fig:somkachinfeatureimportance} for the Uyghur pieces.}
	\label{fig:somuyghurfeatureimportance}
\end{figure}

As mentioned above, each of the four used features of spectral centroid, roughness, sharpness, and loudness have a value for their mean and standard deviation at each neuron position. These are displayed in \textbf{Figure} \ref{fig:somkachinfeatureimportance} including only the Kachin and in \textbf{Figure} \ref{fig:somuyghurfeatureimportance} including only the Uyghur pieces. These plots allow examining the feature vectors responsible for placing pieces at certain neurons.

These plots show full detail, still are harder to read. The up-down axis of each plot shows the strength of each feature normalized with respect to the maximum and minimum value for this feature all over the map. The left-right axis shows the importance of each feature with respect to sorting a musical piece right at this neuron. Although the association of pieces to neurons are performed as the correlation between the neuron and musical piece feature vectors, features similar in both contribute considerably stronger to the musical piece placement on the map. These features are more on the left of each subplot, while features on the right are less nor nearly not important for the placement.

As an example, the neuron at the very bottom on the very right has a very high spectral centroid mean (dark blue) on the very left with the other features separated to the right. Therefore, sorting musical pieces on this neuron is caused by the high spectral centroid mean of these pieces. Another case is the neuron in the fourth row on the very left. Here all features have a medium value (up-down axis), but here most are on the very left. Therefore sorting a musical piece here means that the values of all features on the left fit very well with the neuron features. Yet, a third example is the neuron in the first row in column six. Only one feature, the sharpness standard deviation (light green) is on the left, meaning this is the important feature, still, its value is medium (up-down axis).

\section{Conclusions}

Connections between Kachin and Uyghur pieces are compared with respect to tonal system and timbre. The tonal systems clearly split both ethnic groups, where Uyghur pieces are closer to just intonation while Kachin music shows sharp fifth and fourth, next to a strong sixth. The Uyghur pieces have a strong seventh. The Uyghur pieces are much more similar to one another in terms of the tonal system, while Kachin music shows a greater variety.

Contrary to tonal systems, in terms of timbre, Kachin pieces are more similar to one another compared to Uyghur music. Kachin music also has less standard deviations, especially with respect to sharpness and brightness. One might find this corresponding to subjective findings of the much more expressive music of the Uyghur people and smoother performance with the Kachin.

The Uyghur bel canto pieces form a timbre group on their own, still again distant from Kachin music. They also have a high standard deviation, still mainly on loudness, by at the same time low loudness mean, different from the high standard deviation of sharpness present with the other Uyghur pieces. We can not conclude from the present database that bel canto in general is characterized by these timbre features. Still, the strong loudness changes in the song are clearly corresponding to form parts, large-scale forms, those with and without singing.

Although geographically quite distant, there is no considerable difference between music from Myitkyina and Pota-O. The Wunpawng marching band stands out from all other Kachin music with its highest sharpness and loudness. Still, it does not fit into Uyghur timbre, too, as the loudness shows no considerable standard deviation.

Interesting are the two timbre similarities between traditional and Rock music, the Uyghur pieces of sharpness standard deviation.

The proposed method of using MIR and ML algorithms to compare music from different ethnic groups has four main advantages compared to traditional ethnomusicological analysis 'by hand':

\begin{itemize} 
	\item{The results are objective.} As the algorithms used are based on decades of research on music and sound perception, we can be sure that the differences are musically meaningful, and, therefore, the measures are objective. The used parameters, timbre, pitch, or tonal system are clearly the main features of music. Although additional features could be used (rhythm, melismas, musical instruments, etc.), those investigated are main features, and differences and similarities found here are crucial for the comparison of the music of ethnic groups.
	\item \emph{The method uncovers unexpected differences or similarities.} The near-perfect clustering in terms of tonal systems has not been expected. Although differences could be assumed, the ability of tonal system analysis to clearly identify pieces as having its origin in one of the two groups and that tonal system is the feature separating both most clearly could not be assumed in the first place.
	\item \emph{The results are qualitative.} That Uyghur pieces show enhanced dynamics compared to Kachin music might be clear from listening. Still that this mainly differs in the standard deviation of sharpness, also to some amount in brightness, and not in SPL or roughness, is much harder to determine aurally. As sharpness mainly focuses on 1-3 kHz region, this effect corresponds to the so-called loudness-war of Western popular music, where pushing energy into this frequency region leads to increased perceptual loudness. Links to instrument building of dutars or setars are interesting to draw in the future.
	\item{The results are quantitative.} The differences in tonal systems show main similarities and differences in tonal systems of these two groups quantitatively, showing relevant and irrelevant difference regions. The use of a tonal system much closer to just intonation of the Uyghur compared to the Kachin can be quantified.
	\item \emph{The method allows details.} The presented results aim to find differences in the music of two ethnic groups. Still, many details in the analysis of pieces can additionally be found in the analysis results, again objectively, quantitatively, as well as qualitatively.
\end{itemize}

The analysis might be considered a first step of defining ethnic identity. Still, on the other hand, many similarities have been found too. The study is therefore also about both unity and diversity.

\section{Funding Information}

The project was funded by the Deutsche Forschungsgemeinschaft (DFG) as project 'Music Information Retrieval-Basierte Dateninfrastruktur für Ethnographische Tonträgerarchive',  GZ BA 2208/11-1

\end{document}